\documentstyle[12pt]{article}
\input epsfig.sty
\newcommand{\be}[3]{\begin{equation}  \label{#1#2#3}}     
\newcommand{\ee}{ \end{equation}}
\newcommand{\ba}{\begin{array}}
\newcommand{\ea}{\end{array}}

\renewcommand{\arraystretch}{1.6}

\begin{document}

\thispagestyle{empty}
\rightline{UPR-738-T}
\rightline{HUB-EP-97/16}
\rightline{hep-th/9702205}
\vspace{1truecm}
\centerline{\bf \large
BPS-Saturated Bound States  of  Tilted P-Branes}
\vspace{.4truecm}
\centerline{\bf \large
in  Type II String Theory
}
\vspace{1.2truecm}
\centerline{\bf Klaus Behrndt$^a$ \footnote{e-mail:
 behrndt@qft2.physik.hu-berlin.de}
and Mirjam Cveti{\v c}$^b$ \footnote{e-mail: cvetic@cvetic.hep.upenn.edu}}
\vspace{.5truecm}
\centerline{$^a$ Humboldt-Universit\"at, Institut f\"ur Physik}
\centerline{Invalidenstra\ss e 110, 10115 Berlin, Germany}
\vspace{.3truecm}
\centerline{$^b$ Department of Physics and Astronomy}
\centerline{University of Pennsylvania, Philadelphia, PA 19104-6396}

\vspace{2.2truecm}


\vspace{.5truecm}

\begin{abstract}
We found BPS-saturated solutions of $M$-theory and Type II string
theory which correspond to (non-marginally) bound states of p-branes
intersecting at angles different from $\pi/2$.  These solutions are
obtained by starting with a BPS marginally bound (orthogonally)
intersecting configurations of two p-branes (e.g, two four-branes of
Type II string theory), performing a  boost transformation at an
angle with respect to the world-volume of the configuration,
performing $T$-duality transformation along the boost-direction,
$S$-duality transformation, and $T$- transformations along the
direction perpendicular to the boost transformation. The resulting
configuration is non-marginally bound BPS-saturated solution whose
{\it static metric} possesses the off-diagonal term which {\it cannot}
be removed by a coordinate transformation, and thus signifies an
angle (different from $\pi/2$) between the resulting intersecting
p-branes.  Additional new p-branes   are bound to this
configuration, in order to ensure the
stability of such a static, tilted configuration.
\end{abstract}

\bigskip \bigskip
\newpage

\noindent
{\bf  1. Introduction and procedure} \medskip

\noindent
Recently a number of BPS-saturated  configurations
representing (non-marginally) bound states  of  various p-brane
configurations in $M$- and Type II
string theory  were obtained.\footnote{For a recent extensive discussion
of such configurations see Ref. \cite{tsey} and references therein.}
In general such  configurations are
obtained by  performing a subset of $U$-duality transformations on a 
marginally bound configuration, representing BPS states of (orthogonally
intersecting) p-branes at a threshold, specified by independent harmonic
functions  for
each p-brane \cite{pa/to}. Such solutions, obtained up to this point,  are
interpreted as bound state configurations of a set of  p-branes  whose
pairs
are either {\it parallel} to each-other or intersect {\it orthogonally},
and in
some cases contain a wave along one of the p-brane world-volume direction.
Thus, all the static solutions obtained in this class have {\it
diagonal} internal metric.

\smallskip

The aim of this paper is to find static
non-marginally bound BPS configurations with {\it non-diagonal}
internal metric (which cannot be removed by a coordinate
transformation). Such configurations therefore signify
bound states of p-branes  at {\it angles} different from zero and $\pi/2$.
They are believed to be important when addressing higher
dimensional embeddings of the generating solution for the
four-dimensional BPS-saturated black holes (in toroidally compactified
Type II string). This generating solution is parameterized by {\it
five} charges \cite{cvettsey,cvethull}; the four charges correspond to
the
marginally bound state of four orthogonally intersecting
p-branes (e.g., $2\perp 2\perp 4\perp 4$ in Type IIA  string theory),
while the fifth charge, which renders the solution non-marginally
bound, is suspected to signify a tilting of the intersecting
p-branes.\footnote{From the D-brane world-volume perspective a related
issue was addressed in Ref.\cite{balaleigh}.}

\smallskip

The procedure we employ  makes use of the following features of Type II
string
theory. All the p-brane solutions in ten dimensions are connected
to each other by  discrete  ($U$) duality transformations.
E.g., $T$-duality  transformations  transform
all $D$-p-branes into one another and the $S$-duality converts $D$-branes
into NS-NS sector p-branes (the fundamental string or 5-brane).

\smallskip

On the other hand, a p-brane can be rotated by making first a finite
boost at an angle (to the world-volume of the p-brane), then setting
the charge $Q$ of the original p-brane to zero, and the boost
parameter $\beta$ to infinity, while keeping the product $Q \cdot\cosh^2
\beta$ finite.  A resulting configuration is a fundamental string
along the direction of the original boost.  Again, by performing
discrete $U$-duality transformations we can now create all other types of
p-branes oriented along this direction.

\smallskip

In the case when the boost is taken finite such configurations have an
interpretation as an interpolating, non-marginally bound state. Many
of these (non-marginally) bound states, which are obtained by finite
boosts, are discussed in the literature
\cite{ru/ts,tsey,Papadopetal}.  Note, however, that after
the boost transformation such configurations become stationary
($G_{0m} \neq 0$). By performing
$T$-duality transformations along the boost direction and a direction
perpendicular to the boost, the resulting solution becomes {\it
static}. So the boost parameter does not only create an additional brane,
but
it also rotates the whole configuration.

\smallskip

As the last ingredient needed to obtain a tilted configuration is to
perform a (continuous) $S$-duality ($SL(2,{\bf R})$ transformation
between the two $T$-duality transformations. Such a transformation,
while affecting the nature of the off-diagonal terms in the metric, it
leaves the resulting configuration static,
however, now with a non-diagonal internal metric.
\medskip

We do the above sequence of transformations on a representative example of
two  (orthogonally ) intersecting,
4-branes  ($4\perp 4$) with each of the 4-brane specified by
a harmonic
function.\footnote{We  choose that  type of the starting
configuration for the simplicity of obtaining the final structure of the
metric. Note that the configuration $2\perp 4$  is also of special
interest because it is expected that  the resulting  tilted
configuration of a 2-brane and 4-brane may naturally allow for an
interpretation of a special case
$2\subset 4$\cite{Papadopetal} which preserves $1/2$ of supersymmetry. Note
also, that all such ten-dimensional configurations can be
lifted to eleven-dimensions, thus obtaining  analogous
configurations of $M$-theory.} This is a marginally bound  BPS
configuration which preserves
$1/4$ of supersymmetry.

\smallskip

On this initial configuration we then perform the following  steps:
\begin{itemize}
\item
Perform a boost $\beta$  at an angle $\theta$ to the world-volume of the
intersecting brane.
\item
 Perform  a sequence of $T$-$S$-$T$-duality  transformations\footnote{This
 solution generating technique has been used in different
 context, e.g. in \cite{Bakas}.} , where the
two
discrete $T$-duality transformations are along the boost direction and a
direction
perpendicular to the boost. The (continuous) $S$-duality transformation is
parameterized by an $SO(2)$ angle $\chi$.
\item
As the last step we identify the constituent
 p-branes forming the bound state and the
angles  between them.
\end{itemize}

\begin{figure}[t]
\begin{center}
\mbox{\epsfig{file=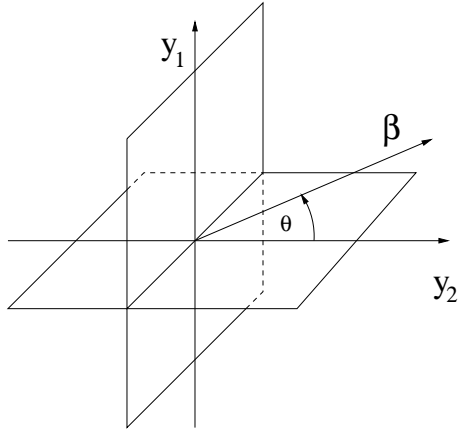, width=60mm}}
\end{center}
\vspace{-6mm}
\caption{From the intersection of two 4-branes we pick out
two world-volume coordinates ($y_1$ and $y_2$). The  boost $\beta$
is made under an angle
$\theta$. We   $T$-dualize this configuration along both  the boost direction and the
direction orthogonal to the  boost, while in between the two $T$-duality
transformations we perform the $S$-duality transformations  with an $SO(2)$
 angle
$\chi$.  
}
\end{figure}

This procedure is described in Figure 1. Few additional comments are
in order. After the first step, we have a stationary solution.
Recall, that in order to obtain a static solution we have to perform
$T$-duality twice along   two orthogonal directions in the
rotation plane. The resulting bound state consists not
only of  two 4-branes, but also contains additional
p-branes. However, such a resulting configuration
 still contains only  perpendicular  and parallel p-branes. Again, in
order to get a configuration where the branes are tilted with respect
to each other
we have to include a continuous $S$- transformation in between. Note also,
that since we started with a Type IIA string solution, after the first
$T$-duality we have a Type IIB string solution, and the $S$-duality
mixes the fundamental string part with a $D$-string part.


\medskip

\bigskip

\noindent
{\bf  2. Explicit transformations} \medskip

\noindent
The starting point is the  (orthogonal) intersection of two 4-branes
(intersecting at a 2-brane),
with  the following  form of the ten-dimensional metric,
the anti-symmetric  (R-R) 4-form field strength and the dilaton:
\be010
\ba{l}
ds_0^2 = {1 \over \sqrt{H_1 H_2}} \left[ (-dt^2 + H_1 dy_1^2 + H_2 dy_2^2)
 + dz_1^2 + dz_2^2 +  H_1 dy_3^2 + H_2 dy_4^2 + H_1 H_2
 d\vec{x} d\vec{x} \right] \\
 F^4 = (\ ^* dH_1 \wedge dy_2 \wedge dy_4 \ +  \  ^* dH_2 \wedge dy_1
 \wedge dy_3) \wedge dt \wedge dz_1 \wedge dz_2 \  \ \\
 e^{-2\phi} = \sqrt{H_1 H_2} \ .
 \ea
\ee
and the harmonic functions $H_{1,2}$   are given by
\be012
H_{1,2} = 1 + {Q_{1,2} \over r} \ .
\ee
with $r^2 = \vec{x} \vec{x}$ (with ${\vec x}=(x_1, x_2 , x_3)$).

\medskip

\noindent
{\em Coordinate transformations (boost  at an angle)}

\medskip

\noindent
We first transform the metric, and as the next step we shall
discuss the R-R fields.
Since we  perform  a boost $\beta$  under an angle $\theta$, we first rotate
the $(y_1, y_2)$ plane and make a boost along the new $y_1$ coordinate.
If we denote the original metric by $G^{(0)}_{rs}$ ($r,s = t , y_1 , y_2$)
the new metric $G_{rs}$ is
\be020
G = (\Omega_R \Omega_B)^T G^{(0)} (\Omega_R \Omega_B) =
 {1 \over \sqrt{H_1 H_2}}\; (\Omega_R \Omega_B)^T \pmatrix{
 -1 && \cr & H_1 & \cr && H_2} (\Omega_R \Omega_B)
\ee
with the rotation and boost transformation given by
\be030
\Omega_R = \pmatrix{ 1 & 0 & 0 \cr 0 & \cos \theta & \sin \theta \cr
 0 & -\sin \theta & \cos \theta} \quad , \quad
\Omega_B = \pmatrix{ \cosh \beta & -\sinh \beta & 0  \cr
 -\sinh \beta & \cosh \beta & 0 \cr 0 & 0 & 1} \ .
\ee
The new metric can be written as
\be040
G_{rs} = { 1 \over \sqrt{H_1 H_2}} \left( \eta_{rs} + {Q_{rs}  \over r}
\right)
\ee
with
\be050
\ba{ll}
Q_{00} = ( Q_1 \cos^2 \theta + Q_2 \sin^2 \theta ) \sinh^2 \beta \ , &
Q_{11} = ( Q_1 \cos^2 \theta + Q_2 \sin^2 \theta ) \cosh^2 \beta \ , \\
Q_{01} = - ( Q_1 \cos^2 \theta + Q_2 \sin^2 \theta )
\sinh \beta \cosh \beta \ ,&
Q_{22} = (Q_1 \sin^2 \theta + Q_2 \cos^2 \theta) \ , \\
Q_{02} = -(Q_1 - Q_2) \sin \theta \cos \theta \sinh \beta  \ , &
Q_{12} = (Q_1 - Q_2) \sin \theta \cos \theta \cosh \beta.
\ea
\ee
The resulting metric is thus stationary with  non-vanishing
$G_{01}$ and $G_{02}$ components.

\medskip

\noindent
{\em Duality transformations}

\medskip

\noindent
By performing $T$-duality twice along the  new $y_1$ and $y_2$
 directions, we obtain a
static metric, but with  additional components of the  anti-symmetric
tensor.\footnote{An explanation of how $T$-duality along the boost  leads to a
fundamental string in the dual picture was given in \cite{cp}.}
Note that the starting configuration did not have
an anti-symmetric tensor, i.e.,  $B_{rs}=
B_{rs}^{(0)}=0$.  The new metric and the anti-symmetric tensor after the
sequence of two
$T$-duality operations  is of the form:
\be060
\hat{G}_{rs} = {1 \over D_0} \pmatrix{ \det G & 0 & 0 \cr
 0 & G_{22} & - G_{12} \cr 0 & -G_{12} & G_{11}} \qquad ,
 \qquad \hat{B}_{rs} = {1 \over D_0} \pmatrix{ 0 & D_1 & -D_2 \cr
 -D_1 & 0 & 0 \cr D_2 & 0 & 0 }
\ee
with the sub-determinants of $G_{rs}$ in (\ref{040}):
$D_0 = G_{11} G_{22} - G_{12}^2$, $D_1 = G_{10} G_{22} - G_{20}G_{12}$,
$D_2 = G_{10} G_{21} - G_{11} G_{20}$.
Using (\ref{020}) we find: $\det G = \det G^{(0)} = - (\sqrt{H_1
H_2})^{-1}$. The non-vanishing time-like components of the anti-symmetric
tensor indicate, that the new configuration contains a fundamental
string.

\medskip

Since we started with a Type IIA string solution, after the first $T$-duality
we have a Type IIB solution.  On this Type IIB solution we
can perform a $SL(2,{\bf R})$ transformation and perform the second
$T$-duality transformation afterwords.
However, the $SL(2,{\bf R})$ transformation mixes
the NS-NS fields with the R-R fields. Thus,  before we
display the final form of the metric and the anti-symmetric tensor, we first
discuss the R-R gauge fields. 

By performing the
rotated boost we obtain for R-R 4-form field strength in (\ref{010})
\be062
\ba{l}
F^4 = \ ^*dH_1 \wedge dy_4 \wedge dz_1 \wedge dz_2 \wedge
 \left( -\sin \theta \,
 dt \wedge dy_1 + \cos \theta(\cosh \beta \, dt - \sinh \beta \, dy_1)
 \wedge dy_2 \right) \\
\qquad + \ ^*dH_2 \wedge dy_3 \wedge dz_1 \wedge dz_2
\wedge \left( \cos \theta \,
 dt \wedge dy_1 + \sin \theta (\cosh \beta \, dt - \sinh \beta \, dy_1)
 \wedge dy_2 \right) \ .
\ea
\ee
Next, after $T$-duality transformation along the $y_1$ direction we obtain
a  R-R-torsion and a 5-form field strength contribution.
Using the  transformations  given in \cite{be/hu/or} we find
\be064
\ba{l}
 H^{R-R}_0 =  \cosh \beta ( \cos \theta \ ^*dH_1 \wedge dy_4 + \sin \theta
 \ ^*dH_2 \wedge dy_3 )\wedge dt \wedge dy_1 \wedge dy_2 \wedge dz_1
 \wedge dz_2 \\
 F^5 = dH_1 \wedge dz_1 \wedge dz_2 \wedge dy_4 \wedge (-\sin\theta dt
 + \cos\theta \sinh\beta dy_2) + \\
 \qquad dH_2 \wedge dz_1 \wedge dz_2 \wedge dy_3 \wedge ( \cos\theta dt
 + \sin\theta \sinh\beta dy_2) + (dual) \ .
 \ea
\ee
Since we started with a configuration of two 4-branes  there is {\it no} R-R
scalar for the Type IIB  string solution.
 As the next step we  perform the $S$-duality transformation, specified by  an
 $SO(2)$ angle $\chi$, which leaves
$F^5$ invariant. The new NS-NS and R-R-torsions are given by
\be066
\pmatrix{H^{NS{\rm -}NS} \cr H^{R{\rm -}R}} = \pmatrix{ \cos\chi \, H^{NS{\rm
-}NS}_0 + \sin\chi
\, H^{R{\rm -}R}_0 \cr -\sin\chi \, H^{NS{\rm -} NS}_0 + \cos\chi \, H^{R{\rm-}R}_0}   ,
\ee
where $H^{NS}_0$ is
\be067
H^{NS{\rm -} NS}_0 = d{G_{12} \over G_{11}} \wedge dy_1 \wedge dy_2 + d {G_{10} \over
G_{11}} \wedge dy_1 \wedge dt \
\ee
with $G_{rs}$ defined in (\ref{040}). Now, as the last step we perform
the  second $T$-duality along  $y_2$ direction.
 Importantly, the R-R part in $H^{NS{\rm -} NS}$ {\it does not contribute to the
 new metric},
since it has no components in the $y_2$ direction.

 Before further addressing the
NS-NS fields we continue the discussion of the R-R fields.
First, there is  a non-trivial vector field part
\be068
F^2 = -\sin\chi \, d{G_{12} \over G_{11}} \wedge dy_1 \ .
\ee
However,  there is no charge associated with this field strength, i.e.
the corresponding
integrals at spatial infinity vanish ($\int F = \int ^*F =0$).
Therefore, there is {\it no}  0- or 6-brane contained in the resulting
 bound state configuration.

The 4-form gauge fields split into an electric and magnetic part.
The electric part is given by
\be069
F^4 = - \sin\chi \, d{D_1 \over D_0} \wedge dy_1 \wedge dy_2 \wedge dt \ .
\ee
Hence, our configuration has to contain a two-brane, which couples to
this field strength. The magnetic part can be written in terms of
components as
\be072
\ba{l}
F^4_{2 \mu\nu\rho} = \cos\chi H^{R-R}_{0 \, \mu\nu\rho} \\
F^4_{\mu\nu\rho\lambda} = F^5_{\mu\nu\rho\lambda 2}
+ \partial_{[\mu} {G_{12} \over G_{11}} B^{R-R}_{0 \, \rho\lambda ]}
\delta_{\nu 1} \ .
\ea
\ee
Therefore, our final configuration is a bound state of two magnetic
4-branes, one electric two-brane ($F^4_{0mnp}$),
 a fundamental string ($B^{NS{\rm -} NS}_{0m}$)
and a $NS$-5-brane (coming from $H_0^{R-R}$ in $H^{NS-NS}$,
see (\ref{066})).  Again there are no 0- or 6-branes.

\medskip

Now we address the  metric, the anti-symmetric tensor and the dilaton field.  After the $T$-$S$-$T$ transformations
the first three
components of the metric, the anti-symmetric tensor and  the
dilaton take  the  final form:

\be070
\ba{llll}
\hat{G}_{00} = \rho {\det G \over D_0} \ , &
\hat{G}_{11} = {\cos^2 \chi \over \rho} {G_{22} \over D_0} +
{\sin^2 \chi \over \rho } e^{-2 \phi} \ , &
\hat{G}_{22} = {G_{11} \over \rho D_0} \ , &
\hat{G}_{12} = - {\cos \chi \over \rho} {G_{12} \over D_0}\ , \\
 \hat{B}_{10} = \cos \chi \, {D_1 \over D_0} \ , &
 \hat{B}_{20} = - \cos\chi \, {D_2 \over D_0} \   , &
  e^{-2 \hat{\phi}}= e^{-2\phi}{D_0 \over \rho^3}  
 &
\ea
\ee
with: $\rho^2 = \cos^2 \chi + \sin^2 \chi \, e^{-2\phi} G_{11}$
and $\chi$ is an angle of $SO(2) \subset SL(2,{\bf R})$.
Inserting our metric and dilaton, $\rho^2$ defines
a new harmonic function
\be080
\hat{H} \equiv \rho^2 = 1 + { Q_{11} \sin^2 \chi \over r} \ .
\ee
Also, using the metric (\ref{040}) we can define two further
harmonic functions $\tilde{H}_{1,2}$ by
\be090
D_0 = {\tilde{H}_1 \tilde{H}_2 \over H_1 H_2} \ .
\ee
with the charges $\tilde{Q}_1$ and $\tilde{Q}_2$.
Note that  $\tilde{Q}_1 \tilde{Q}_2 = Q_1 Q_2 \cosh^2 \beta$, thus,
 $\tilde{H}_{1,2}=1$ if $H_{1,2}=1$.

\smallskip

Note,  the metric  (\ref{070}) of the final configuration is {\it static} with
 the {\it off-diagonal} metric component which cannot be removed by a coordinate
 transformation, thus signifying the angle between constituent  p-branes,
 different from $\pi/2$. Note also that non-zero components of the
 two-from field  (\ref{070}) signal
 the existence of the {\it fundamental} string.

 Before discussing further the features of the  resulting configuration we
 turn now to  a discussion of a  special cases.

\bigskip

\noindent
{\bf  3. A single four-brane bound state} \medskip

\noindent
We first  consider the case of a single 4-brane. Thus we take the limit:
$Q_2=0$, $Q_1 = Q$. As consequence $\tilde{H}_2 =1$  and
we obtain for the metric
\be100
\ba{rcl}
d\hat{s}^2&= {\sqrt{H} \over \tilde{H} \sqrt{\hat{H}}}&\left[ - \hat{H} dt^2
 + ((1+ {Q_{22} \over r}) \cos^2 \chi + \tilde{H} \sin^2 \chi) dy^2_1 +
 (1+ {Q_{11} \over r}) dy^2_2 - \right. \\
 && \left. - 2 {Q_{12} \over r} \cos \chi \, dy_1 dy_2 \right]
 + \sqrt{\hat{H}} d\bar{s}^2   \\
 &=  {\sqrt{H} \over \tilde{H} \sqrt{\hat{H}}}&\left[ - \hat{H} dt^2
 + dy_1^2 + dy_2^2 + {1 \over r} \left( (Q_{22} \cos^2 \chi +
 \tilde{Q} \sin^2 \chi ) dy_1^2 + Q_{11} dy_2^2 - \right. \right. \\
 && \left. \left.  -
 2 Q_{12} \cos \chi \, dy_1 dy_2 \right) \right] +  \sqrt{\hat{H}} d\bar{s}^2
\ea
\ee
where the $d\bar{s}^2$ denotes the unchanged part of the original metric
(\ref{010}). This metric can be diagonalized by a further rotation
\be110
 \pmatrix{ y_1 \cr y_2 } \rightarrow
 \pmatrix{ \tilde{y}_1 \cr \tilde{y}_2 } =
 \pmatrix{
 \cos \psi & \sin \psi \cr - \sin \psi & \cos \psi}
 \pmatrix{ y_1 \cr y_2 } \ .
\ee
As result we get
\be120
 d\hat{s}^2 = \sqrt{H \over \hat{H}} \left[ {\hat{H} \over \tilde{H}}
 (-dt^2 + d\tilde{y}^2_2 ) + d \tilde{y}_1^2 \right] +
 \sqrt{\hat{H}} d\bar{s}^2 \ ,
\ee
and for the angle $\psi$ we find
\be130
\tan \psi = \pmatrix{-{Q_{22}  \over Q_{12}\cos \chi}}_{Q_2 =0}
= - {1 \over \cot\theta \, \cosh \beta \,  \cos \chi} \ .
\ee
There is a second solution given by the angle $\psi + {\pi \over 2}$.
The harmonic functions are
\be140
H= 1 + {Q \over r} \quad , \quad
\tilde{H} = 1 + {{Q(\sinh^2 \beta \;\cos^2 \theta +1)} \over r}
 \quad , \quad
 \hat{H} = 1 + {Q \sin^2 \chi \; \cosh^2 \beta \; \cos^2 \theta \over r}
\ee
(recall: $\theta$ was the angle of the boost, $\beta$ is the boost parameter
and $\chi$ is the  $SO(2)\subset SL(2,{\bf R})$ parameter).
From the discussion of the
gauge fields we know already, that this configuration contains
the following branes
\be150
\ba{lll}
\mbox{4-brane}: & \hat{H} =1\ , \  \tilde{H} =H  \ ,
 & (\beta = \theta = \chi =0) \\
\mbox{fundam.\ string}: & H=\hat{H} =1 \ ,
 &(Q \rightarrow 0 , \beta \rightarrow \infty ; \ \sin \chi = 0) \\
\mbox{2-brane}: & H=1 \ , \ \hat{H} = \tilde{H}\ , \ &(Q \rightarrow 0 ,
 \beta \rightarrow \infty ; \ \cos \chi = 0) \  \\
 \mbox{NS-5-brane}: & H= \hat H = \tilde H \ , & (\beta \rightarrow 0 \ ,
\sin\chi \cos \theta =1)  \ .
\ea
\ee
The solution (\ref{120}) is therefore a bound state of these objects 
($4+2+ 1_f+5_{NS}$)  and it breaks 1/2 of supersymmetry.  For the special
cases (\ref{150}) the angle $\psi$ is trivial; $\psi = 0, {\pi \over
2}$.  But if all branes are turned on the whole configuration is
rotated by the angle $\psi$ with respect to the original location,
$(y_1, y_2)$ vs.\ $(\tilde{y}_1, \tilde{y}_2)$.  However, the
constituents of (\ref{120}) are still perpendicular to each other.

\bigskip

\noindent
{\bf  4. The $4 \times 4$ bound state} \medskip

\noindent
Finally we return to the general solution obtained from
considering {\it two} intersecting 4-branes (\ref{010}),
i.e.\ a configuration that breaks $1/4$ of supersymmetry.
In this case the $T$-$S$-$T$ transformed metric is of the form (\ref{070}) and
can be cast in the form:
\be160
\ba{rl}
d\hat{s}^2 = {\sqrt{ H_1 H_2} \over \tilde{H}_1 \tilde{H}_2
\sqrt{\hat{H}}}&\left[ - \hat{H} dt^2 + \left( (1+ {Q_{22} \over r})
\cos^2 \chi + \tilde{H}_1 \tilde{H}_2  \sin^2 \chi \right)) dy^2_1 +
 (1+ {Q_{11} \over r}) dy^2_2 - \right. \\
 & \left. - 2 {Q_{12} \over r} \cos \chi \, dy_1 dy_2 \right]
 + \sqrt{\hat{H}} d\bar{s}^2 \ .
\ea
\ee
In comparison to the former case (\ref{100}) there is now a main
difference. Before, we could factorize the dependence on the
radius $r$ (second eq.\ in (\ref{100})) so that by a rotation we
could get rid of the off-diagonal part. However, now this is not
possible, because of the $\tilde{H}_1 \tilde{H}_2$ part in (\ref{160}).
Only for the special cases that one of the  harmonic functions
$\tilde{H}_{1, 2}$ is trivial, i.e. $\tilde{Q}_1 \tilde{Q}_2 =
Q_1Q_2 \cosh^2\beta =0$,  or if
 $\sin \chi \, \cos \chi =0$ we still have a static
configuration. E.g.\ for $\cos \chi =0$ we get
\be170
 d\hat{s}^2 = \sqrt{H_1 H_2 \over \hat{H}} \left[ {\hat{H} \over
\tilde{H}_1 \tilde{H}_2} (-dt^2 + dy^2_2 ) + dy_1^2 \right] +
 \sqrt{\hat{H}} d\bar{s}^2
\ee
which is a bound state of  $4 \times 4 + 2 + 5_{NS}$.
 Note, that $\cos \chi =0$
turns off the fundamental string, i.e. $B_{0r}=0$ (see (\ref{070})).
Again, in this case all objects are orthogonal. Since the first part is
completely symmetric with respect to $Q_1 \leftrightarrow Q_2$,
both 4-branes are now parallel in the $(y_1,y_2)$ directions, but
still orthogonal in the $(y_3, y_4)$ directions, see (\ref{010}).
So we have rotated both 4-branes into each other, at least
partly.\footnote{Note however, that this configuration still preserves only
$1/4$ of supersymmetry.}

In the case $Q_2=0$ we got the solution (\ref{120}). On the other hand,
if $Q_1 =0$ we find the solution with the same structure
\be180
 d\hat{s}^2 = \sqrt{H \over \hat{H}} \left[ {\hat{H} \over \tilde{H}}
 (-dt^2 + d\tilde{\tilde{y}}^2_2 ) + d \tilde{\tilde{y}}_1^2 \right] +
 \sqrt{\hat{H}} d\bar{s}^2
\ee
but now with a new angle
\be190
 \tan\psi' = \pmatrix{{Q_{11} \cos \chi \over Q_{12}}}_{Q_1=0} =
-\tan \theta \, \cosh \beta \, \cos \chi \ .
\ee
As before there is a second angle with an additional rotation of $\pi
\over2$.

\smallskip

How about the general case with  charges $Q_1$ and $Q_2$ non-zero? The location of the two 4-branes can be determined by
setting one of the charges to zero. We saw, that our procedure rotated
both 4-branes, one by $\psi$ and the other by $\psi'$. As a result, both
4-branes are now under an angle: ${\pi \over 2} - \psi + \psi'$ (see
Figure 2).  This tilted state is however only stable due to
additional p-branes, in general a 2-brane, a fundamental string and
an NS-5-brane. But by setting $\beta=0$ ($D_1 =D_2 = 0$)
we can turn off the 2-brane and the fundamental string and keep only
the NS-5-brane. Note, the electric charges in our
configuration are a consequence of the boost transformation. 
In special limits (typically if $\sin
\chi \, \cos \chi =0$ or $\beta \rightarrow \infty$) we recover an
orthogonal configuration.

There is yet another way of interpreting  our configuration. The $y^1$ and
$y^2$ direction defines a deformed torus. Compactifying  along these
 coordinates,
yields a complex scalar field $U$ which parameterizes the two-torus. The
off-diagonal term of the metric has the consequence that $U$ contains an
imaginary-axion part. This axion part is also responsible that the
two circles of
the torus do not intersect orthogonally. So, the branes that are wrapped
around these two circles intersect now at an angle different from $\pi/2$.
By the $T$-$S$-$T$
operation we had mapped a non-trivial $S$-modulus on the Type IIB side to
an non-trivial $U$ modulus (=tilting of the internal space) on the Type
IIA side. Thus from the duality point of view the tilting of the branes
was the expected result.

\begin{figure}[t]
\begin{center}
\mbox{\epsfig{file=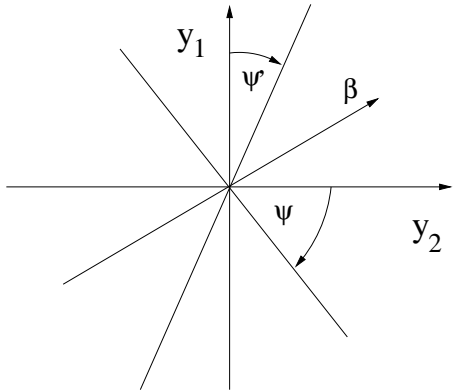, width=60mm}}
\end{center}
\vspace{-6mm}
\caption{After making a finite boost between both four-branes
and performing $T$-$S$-$T$-duality, the two 4-branes are tilted with respect to
 each other.
}
\end{figure}

\bigskip \bigskip
\pagebreak

\noindent
{\bf  5. Conclusions} \medskip

\noindent
In this paper we considered an intersection of two four-branes and made a
finite boost between the world-volumes of the branes (see Figure 1). The
resulting configuration has a stationary metric ($G_{0m} \neq 0$).  By
$T$-dualizing these off-diagonal directions, the metric becomes
static, but {\em not} diagonal. This new configuration is a
(non-marginal) bound state of two four-branes, a two-brane,  a
fundamental string and an NS-5-brane
($4 \times 4 + 2 + 1_f + 5_{NS}$)  , which has $1/4$ of
unbroken supersymmetry.  In the single four-brane limit (see Section 3),
 it is possible to diagonalize the metric. In this case our
procedure rotated the original four-brane  and added a membrane, 
a fundamental string and an NS-5-brane. 
However, for the $4 \times 4$ case (see
Section 4), this is  not possible. By comparing the positions of the new
four-branes, we find that generically they are not orthogonal to each
other. Both branes have been rotated  but with a
different angle and therefore they intersect each other at an angle different
from $\pi/2$.  From the angles in (\ref{130}) and (\ref{190})
we see, that both have the same sign and thus both branes are
rotated in the same direction. Note, also that for arriving at this
conclusion it
was essential that we made a $S$-duality between both $T$-duality
transformations. Had the $S$-duality not been included, the resulting
configuration  would  still have been orthogonal.

The case of the two instersecting four-branes is only an example and
the procedure described in this paper is quite general.
It  can be straightforwardly  adapted \cite{be/cv} to other intersecting brane
configurations  with the qualitatively
same result, that the  (static) branes  are rotated by different angles.

\bigskip

\bigskip

\noindent
{\bf Acknowledgements} \medskip \newline
We would like to thank A. A. Tseytlin for many helpful discussions
and suggestions, and M. Costa for comments.  The work is supported by U.S. DOE Grant
Nos. DOE-EY-76-02-3071 (M.C.), the National Science Foundation Career
Advancement Award No. PHY95-12732 (M.C.), the NATO collaborative
research grant CGR No. 940870 (M.C.) and the Deutsche
Forschungsgemeinschaft (K.B.). K.B.\ thanks the
University of Pennsylvania and
M.C.\ the Humboldt University for the hospitality during visits
where this work was initiated.

\renewcommand{\arraystretch}{1}




\begin{thebibliography}{aaa}
\bibitem{tsey}{A.A. Tseytlin, {\em Composite BPS configurations of p-branes in
 10 dimensions and 11 dimensions}, {\tt hep-th/9702163}.}
\bibitem{pa/to}
 G. Papadopoulos and P.K. Townsend, {\em Intersecting M-branes},
 {\tt hep-th/9603087};
 A.A. Tseytlin, { \em Harmonic superpositions of M-branes}
 {\tt hep-th/9604035}; M. Cveti\v c and A. Sen, unpublished;
 K. Behrndt, E. Bergshoeff and B. Janssen,
 {\em Intersecting d-branes in ten-dimensions and six-dimensions},
 {\tt hep-th/9604168};
 J.P. Gauntlett, D.A. Kastor and J. Traschen,
 {\em Overlapping branes in M-theorie},
 {\tt hep-th/9604179};
 V. Balasubramanian and F. Larsen,
 {\em On D-branes and black holes in 4 dimensions}
 {\tt hep-th/9604189}.\bibitem{cvettsey}{M. Cveti\v c and A.A. Tseytlin, 
{\em Solitonic strings and BPS saturated dyonic black holes},
{\tt hep-th/9512031}.}
\bibitem{cvethull}{M. Cveti\v c and C.M. Hull,
{\em Black holes and U duality},
{\tt hep-th/9606193}.}
\bibitem{balaleigh}{ M. Berkooz, M.R. Douglas and R.G. Leigh
 {\em Branes intersecting at angles}; {\tt hep-th/9606139}, V. Balasubramanian and  R. Leigh, 
 {\em D-branes, moduli and supersymmetry},
 {\tt hep-th/9611165}}.
\bibitem{ru/ts}
 J.G. Russo and A.A. Tseytlin, {\em Waves, boosted branes and BPS states
 in M-theory}, {\tt  hep-th/9611047};
 J.C. Breckenridge, G. Michaud and R.C. Myers,
 {\em More D-brane bound states},
 {\tt hep-th/9611174};
M.S. Costa, {\em Composite $M$-branes}, {\tt hep-th/9609181}.
\bibitem{Papadopetal}
 J.M. Izquierdo, N.D. Lambert, G. Papadopoulos and P.K. Townsend,
 {\em Dyonic membranes}, {\tt hep-th/9508177}.
\bibitem{Bakas}
I. Bakas,
{\em Space-time interpretation of S duality and supersymmetry 
violations of T duality}, {\tt hep-th/9410104}.
\bibitem{cp}
M.S. Costa and  G. Papadopoulos, {\em Superstring dualities and p-brane bound
states}, {\tt hep-th/9612204}.
\bibitem{be/hu/or}
 E. Bergshoeff, C. Hull and T. Ortin, {\em Duality in type II
 superstring effective action}, {\tt hep-th/9504081};
 E. Bergshoeff and M. De Roo, { \em $D$-branes and $T$-duality},
 {\tt hep-th/9603123}.
\bibitem{be/cv}
M. Cveti{\v c} and K. Behrndt, in preparation.
\end{thebibliography}
\end{document}